\documentclass[aps,pra,twocolumn,nopacs,lettersize,superscriptaddress, floatfix]{revtex4-1}
\usepackage{array}
\usepackage{rotating}
\usepackage{multirow}
\usepackage{placeins}
\usepackage{xcolor}
\usepackage{hyperref}
\usepackage{lipsum} 
\hypersetup{
    colorlinks=true,
    linkcolor=blue,
    citecolor=blue,
    citebordercolor=white,
    linkbordercolor=white,
}
\usepackage{graphicx}
\usepackage{float}
\usepackage{amsmath,amssymb,amsfonts}
\usepackage{cleveref}
\usepackage{mathrsfs}

\usepackage{natbib}
\usepackage{tabularx}
\newcolumntype{Y}{>{\centering\arraybackslash}X}
\usepackage[normalem]{ulem}

\newcommand{\eqnreft}[1]{{Eq.~(\ref{#1})}}

\begin{document}
\title{Vortex matter simulation of the one component plasma}

\author{Tyler W. Neely}
\affiliation{Australian Research Council Centre of Excellence for Engineered Quantum Systems, School of Mathematics and Physics, University of Queensland, St. Lucia, QLD 4072, Australia.}
\author{Guillaume Gauthier}
\affiliation{Australian Research Council Centre of Excellence for Engineered Quantum Systems, School of Mathematics and Physics, University of Queensland, St. Lucia, QLD 4072, Australia.}
\author{Charles Glasspool}
\affiliation{Australian Research Council Centre of Excellence for Future Low-Energy Electronics Technologies, School of Mathematics and Physics, University of Queensland, St. Lucia, QLD 4072, Australia.}
\author{Matthew J. Davis}
\affiliation{Australian Research Council Centre of Excellence for Engineered Quantum Systems, School of Mathematics and Physics, University of Queensland, St. Lucia, QLD 4072, Australia.}
\affiliation{Australian Research Council Centre of Excellence for Future Low-Energy Electronics Technologies, School of Mathematics and Physics, University of Queensland, St. Lucia, QLD 4072, Australia.}
\author{Matthew T. Reeves}
\affiliation{Australian Research Council Centre of Excellence for Future Low-Energy Electronics Technologies, School of Mathematics and Physics, University of Queensland, St. Lucia, QLD 4072, Australia.}
\date{\today}

\begin{abstract}
The two-dimensional one-component plasma (OCP) is a foundational model of the statistical mechanics of interacting particles, describing phenomena common to astrophysics~\cite{baiko2021abinitio}, turbulence~\cite{onsager1949statistical}, and the Fractional Quantum Hall Effect (FQHE)~\cite{laughlin1983anomalous}.  Despite an extensive literature~\cite{cardoso2020boundary,baus1980statistical} the phase diagram of the two-dimensional (2D) OCP is still a subject of some controversy~\cite{nosenko20092d,kapfer2015two}.  Here we develop a ``vortex matter"  simulator to experimentally realize the logarithmic-interaction OCP by exploiting the topological character of  quantized vortices in a thin Bose-Einstein condensate superfluid layer. Precision optical-tweezer control of the location of quantized vortices enables direct preparation of the vortex-analog OCP ground state with optional defects, and subsequent heating of the vortex matter from acoustic excitations results in the melting of the  Wigner crystal to the liquid phase.  Our theoretical analysis is in quantitative agreement with experimental observations and demonstrates how effective equilibrium states are achieved through the nonequilibrium dynamics. The vortex matter simulator provides a potential route towards probing a number of open problems in systems with long-range interactions. At equilibrium it could distinguish between the competing scenarios of grain boundary melting~\cite{gasser2010melting} and Kosterlitz, Thouless, Halperin, Nelson, and Young (KTHNY) theory that predicts an intermediate hexatic phase~\cite{nosenko20092d,ryzhov2017berezinskii}. Dynamical simulators could test the existence of predicted edge-wave solitons~\cite{bogatskiy2019edge} which form a hydrodynamic analog of topological edge states in the FQHE. The platform also allows a precise measurement of the superfluid-thermal cloud mutual friction and heating coefficients, and thus could be used as a stringent experimental probe to validate finite temperature theories  in more complex quantum fluid systems.
\end{abstract}

\maketitle

The one-component plasma (OCP) model describes a set of $N$ identically charged, massless particles interacting via the Coulomb potential. In two dimensions it is described by the Hamiltonian ~\footnote{Equation~(\ref{eqn:PVHamiltonian}) corresponds to the case of a free-space geometry. Periodic or spherical geometries do not require the confining potential.}

\begin{gather}   
   \mathcal{H} = H - \Omega M  \equiv - \sum_{j\neq k} \textrm{ln} \left | 	  z_j -z_k \right| 
         +\Omega \sum_j |z_j|^2,   \label{eqn:PVHamiltonian}
\end{gather} 
where $z_j = x_j+i y_j$ defines the particle positions in the $x$-$y$ plane. The particles are embedded in a rigid background of charge density $-\Omega$, which maintains overall charge neutrality and forms an effective confining potential preventing them from escaping 
to infinity.  The equilibria of the 2D OCP can be completely specified in terms of the plasma parameter 
$\Gamma  = q^2/(k_B T)$, where $q$ is the particle charge.  

 \begin{figure}[t!]
    \centering
\includegraphics[width = \columnwidth]{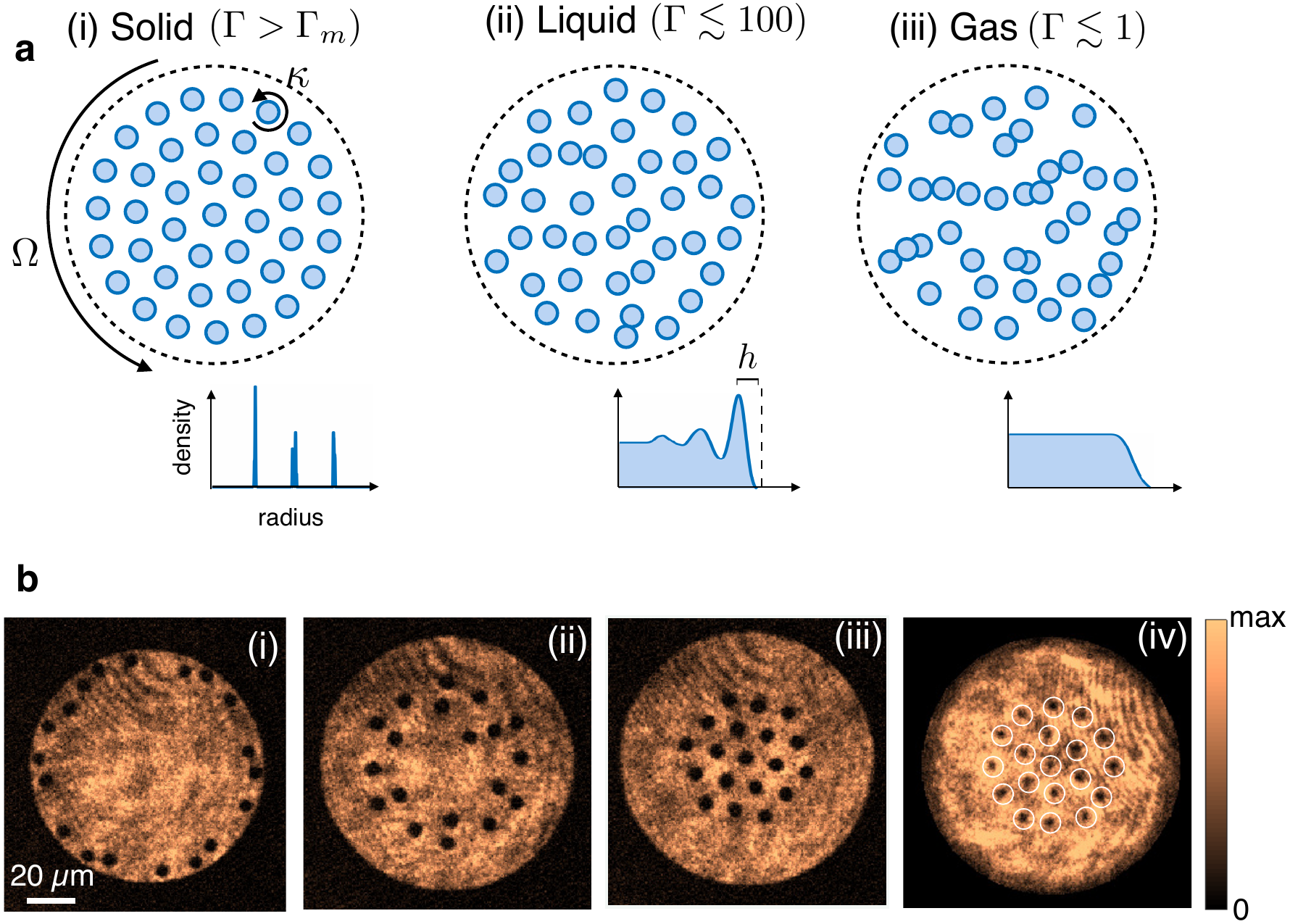}
    \caption{
    \textbf{a} Schematic of the equilibrium states of the OCP. (i) At low temperature (large $\Gamma$) the system forms a (solid) Wigner crystal. (ii) The crystal melts for $\Gamma \sim 100$ forming a correlated liquid phase. The salient features of the strongly-correlated liquid phase are apparent in the radial distribution at intermediate temperature (lower plot) and include: (1) excess density on the edge of the cluster~\cite{bogatskiy2019edge} (2) crystallization at the edge~\cite{cardoso2020boundary}; and (3) spatial compression of the cluster edge by a factor $h$, as compared to a uniform vorticity patch with the same total vorticity (dashed outer circle)~\cite{bogatskiy2019edge,cardoso2020boundary}. (iii) For $\Gamma \sim 1$, the system is in a gas-like phase. 
    \textbf{b} Experimental sequence demonstrating the creation of the Wigner crystal via the use of optical vortex tweezers (i-iii), and observation after a short time-of-flight (iv). White circles mark vortex positions. 
}
    \label{fig:schematic}
\end{figure}

\begin{figure*}[!t]
    \centering
    \includegraphics[width = 0.95\textwidth]{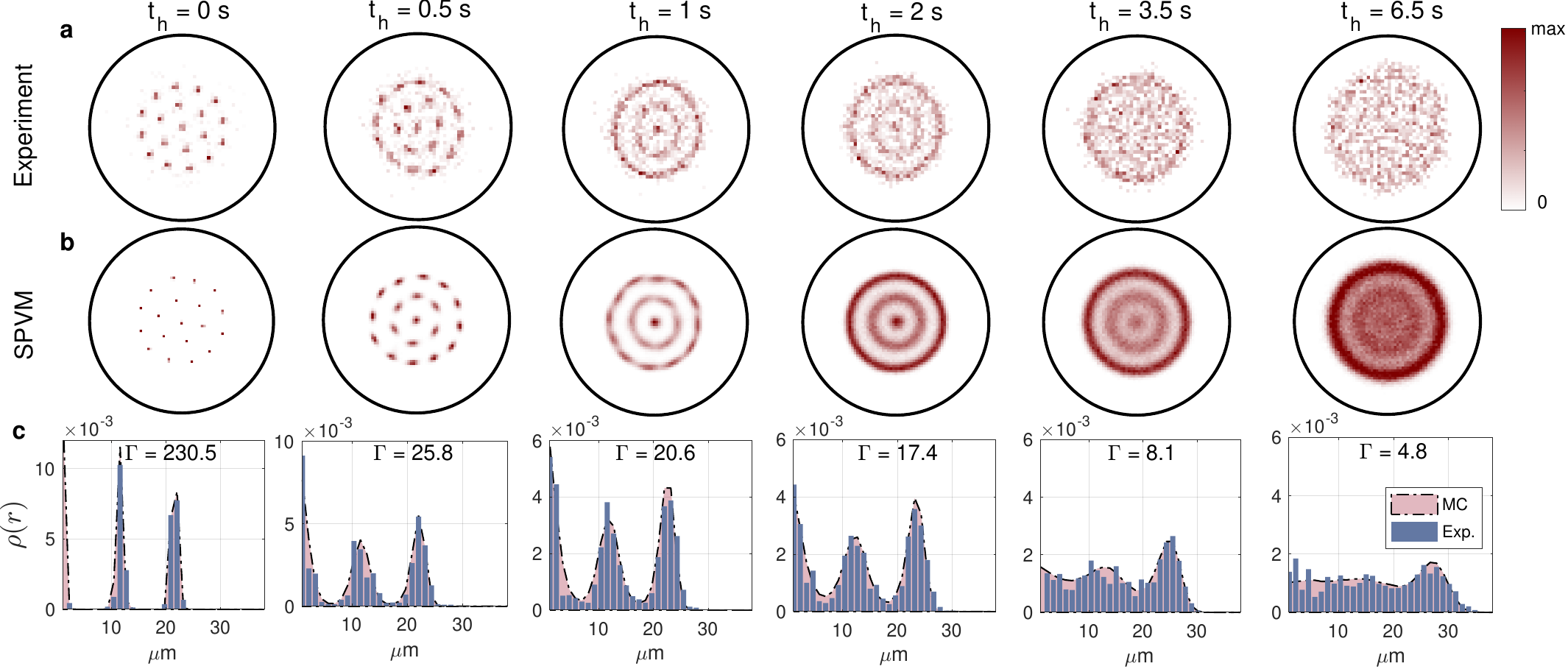}
    \caption{ \textbf{a} \textit{Top row:} experimental vortex position histograms, columns indicate increasing hold time $t_h$. Each histogram contains $\sim 2300$ vortices from $\sim 100$ experimental realizations. At $t_h = 0$~s the vortices form a regular lattice since the initial positions are fixed in place in the laboratory frame by the pinning beams. In subsequent histograms, the vortices undergo free evolution exhibiting melting of the lattice structure with increasing hold time. \textbf{b} \textit{Middle row:} $10^4$ simulations of SPVM dynamics with $\alpha = 3.25\times10^{-3}$ and $\eta = 1.44\times10^{-3}~\xi^2/t_{\xi}$. \textbf{c} \textit{Bottom row:} Radial density $\rho(r)$. Experimental data is represented as bars, with shaded regions indicating Monte Carlo best fits, and inset numbers best-fit values of the plasma parameter $\Gamma$.}
    \label{fig:lattice_histograms}
\end{figure*}

At low temperatures ($\Gamma > \Gamma_m$) the OCP forms a solid crystalline structure known as the Wigner crystal [Fig.~\ref{fig:schematic}\textbf{a}(i)]. It is known that a melting transition occurs at $\Gamma_m \sim 140$~\cite{cardoso2020boundary};  this strongly correlated phase forms a theoretical foundation for real matter under extreme conditions, such as dense stellar matter~\cite{dewitt1987strongly, zerah1992thomas,chamel2008physics} and the interiors of massive planets~\cite{saumon2004shock}. The exact nature of the melting transition is still a subject of study~\cite{nosenko20092d} and has been found to be nonuniversal depending on the type of interactions~\cite{kapfer2015two}. Contradictory results have been found for long-range interactions, with some studies finding evidence for an intermediate hexatic phase predicted by the KTHNY theory~\cite{nosenko20092d,guillamon2009direct,roy2019melting,ryzhov2017berezinskii}, and others not~\cite{Franz1994,Zahn1999,Kapfer2015}. At intermediate temperatures ($\Gamma \lesssim 100$) [Fig.~\ref{fig:schematic}\textbf{a}(ii)], the OCP forms a strongly correlated liquid~\cite{bogatskiy2019edge,cardoso2020boundary,nosenko20092d,alastuey1981classical}, which through Laughlin's plasma analogy~\cite{laughlin1983anomalous}
connects the electron density to the fractional quantum Hall effect (FQHE)~\footnote{For values of $\Gamma = 2/\nu$, the density of electrons in the FQHE for fractional filling $\nu$ relate directly to the  OCP Boltzmann weights $e^{-\Gamma \mathcal{H}}$~\cite{laughlin1983anomalous}).}. Finally, at high temperatures $(\Gamma \lesssim 1)$  [Fig.~\ref{fig:schematic}\textbf{a}(iii)], the OCP crosses over  to a gaseous phase, which has relevance to turbulence, and has recently been explored in e.g.,~\cite{gauthier2019giant,johnstone2019evolution,sachkou2019coherent,reeves2022turbulent}. 

Equation~(\ref{eqn:PVHamiltonian}) directly maps to the Hamiltonian for $N$ vortices in an incompressible and inviscid fluid~\cite{onsager1949statistical} and can thus be realized in a thin superfluid layer. This supports vortices with quantized circulation $\kappa = \pm h /m$, where $h$ is Planck's constant and $m$ is the mass of a superfluid particle. Unlike an ordinary fluid, where the circulation may take on any value, in a superfluid the quantization of the circulation results in its topological protection preventing its decay; this endows the vortices with a particle-like character, with charge $q=\pm 1$ corresponding to the circulation. In the vortex analog, $H = -\sum_{i\neq j} \ln |z_i - z_j|$ stems from the kinetic energy of the fluid, while the effective confining potential arises from rotation of the system  [see Fig.~\ref{fig:schematic}].  The quantity  $M = \sum_j |z_j|^2$, is related to the angular momentum of the superfluid [$L_z\propto ( 1-M)]$, and is a constant of the motion. The  ground state of the vortex matter is the Wigner crystal, which is stationary in a frame rotating at angular frequency $+\Omega$. 

Our experiment utilizes a disk-shaped quasi-2D $^{87}$Rb Bose-Einstein condensate (BEC) as the superfluid confined in a hard-walled optical trap.  The vortices are created, pinned, and positioned in the superfluid through the motion of optical tweezers for each vortex as shown in Fig. 1\textbf{b}(i-iii), (for details, see Methods).  The vortices are initially placed in a near perfect Wigner crystal, before being released by removing the pins. The system then freely evolves in the disk trap, and we follow the evolution of the crystal for increasing hold time $t_h$. 

Due to the destructive nature of imaging the superfluid each experiment results in one set of data.  At intervals of $0.5~$s $- 1.0$~s we observe the location of the vortices, and combine the data over $\sim$100 different runs to form a histogram of vortex positions. Due to background heating of the vortex matter the crystal structure is slowly lost over the 6.5~s duration of the experiment, visible in both the 2D histograms (Fig.~\ref{fig:lattice_histograms}\textbf{a}), and in the radial distribution $\rho(r)$ (Fig.~\ref{fig:lattice_histograms}\textbf{c}).   After $t_h = 3.5$~s, we note that most of the internal structure of the lattice is lost, with a clear observation of excess density on the edge of the cluster. The observed vortex distributions show good agreement with using the equilibrium Metropolis algorithm (Fig.~\ref{fig:lattice_histograms}\textbf{c}) and enable fitting of the plasma parameter $\Gamma$. This suggests that the heating of the crystal is a gradual, quasi-equilibrium melting process. The plasma parameter for the vortex matter system is initially $\Gamma \approx 230$~\footnote{The initial value of $\Gamma$ results from post-selection of the data for this time step, resulting in $\sim 71\%$ of the data runs being analyzed for the MC histogram fit at $t=0$. See Methods for additional details.} consistent with the solid phase ($\Gamma > \Gamma_m \sim 140$), and by the end of the experiment we find $\Gamma \approx 4.8$, indicating that the system is still well within the liquid-like state of the phase diagram at the end of the experiment. We now turn to understanding our observations.

 \emph{Dynamical Origin of the Melting ---} The melting process can be understood by  considering the dissipative processes influencing the vortex dynamics. We have demonstrated that the dynamics of quantized vortices in the atomic gas superfluid are well described by the stochastic point vortex model (SPVM)~\cite{reeves2022turbulent,mehdi2023mutual}:
 \begin{equation}
      \mathrm{d}  z_j = (1 - i \alpha )v_j \,\mathrm{d} t + \sqrt{2 \eta}\, \mathrm{d} W_j.
      \label{eqn:StochasticPV}
 \end{equation}
 Here $z_j = x_j + iy_j$ is the complex-valued coordinate, $v_j$ is the velocity of the $j$th vortex under Hamiltonian evolution, ($v_j \equiv  - i \partial H /\partial z_j^*$),  $\alpha$ is the mutual friction coefficient, and $\eta$ is the vortex diffusion rate~\footnote{Mutual friction typically includes a coefficient $\alpha'$ which opposes the vortex motion, resulting in $\Omega(0)' = (1-\alpha')\Omega(0)$. We find that our analysis is consistent with $\alpha'\sim 0$ so omit this term from Eq.~(\ref{eqn:StochasticPV})}. 
 The real part of the first term results from the Hamiltonian dynamics governed by Eq.~(\ref{eqn:PVHamiltonian}), through
which the system explores the phase space manifold defined by the fixed values \{$H,M,N\}$. The imaginary part of the first term causes like-signed vortices to drift apart (thus lowering the energy $H$) at a rate proportional to their velocities and the friction coefficient $\alpha$.  The second term contains the Brownian-motion Wiener noise processes $\mathrm{d} W_j$, which satisfy $\langle \mathrm{d}W_j^*(t) \mathrm{d}W_k(t') \rangle = \delta_{jk}\delta(t-t') \mathrm{d}t$, with all other correlations vanishing. 

Friction causes an expansion of the vortex matter, leading to loss of energy and angular momentum. While naively this appears at odds with the observed melting, which requires an increase in energy, we note the dynamics governed by Eq.~(\ref{eqn:StochasticPV}), and the statistical mechanics, governed by Eq.~(\ref{eqn:PVHamiltonian}), are invariant under the dilation $\{z,v,t\} \rightarrow \{\lambda z,\lambda^{-1}v,\lambda^2 t \}$. To understand the observed melting, the global expansion associated with $\lambda$ must therefore be scaled out. For the parameters of the experiment~\cite{reeves2022turbulent} we find the expansion is strongly dominated by the friction term. Qualitatively, this is because (due to the net rotation), $v_j$ points azimuthally, and the friction $\propto -i v_j$ thus acts radially outward for 
each vortex. Conversely, the noise may push vortices both radially inward or outward, only  weakly contributing to the expansion. As shown in the Methods, the expansion due to friction acting on the Wigner crystal can be calculated exactly~\cite{supp} and is given by 
 \begin{equation}
\lambda(t) = [1 + 2\Omega(0)\alpha t]^{1/2},
\end{equation}
where $\Omega(0)$ is the rotation rate of the vortex crystal at $t=0$. This gives   $M(t) = \lambda(t)^2M(0)$ and $H(t) = H(0) - N(N-1)\textrm{ln}\,\lambda(t)$. Therefore the expansion may be scaled out by working in the rescaled coordinates $\zeta_j(t) = z_j(t)/\lambda(t)$ which describe the internal structure of the vortex matter relevant to its thermodynamic properties. When the dynamics are viewed in the rescaled coordinates $\zeta_j$, the friction serves to cool the vortex configuration back towards the Wigner crystal~\cite{stockdale2020universal}, while the diffusion causes heating. 

\begin{figure}
    \centering
    \includegraphics[width = \columnwidth]{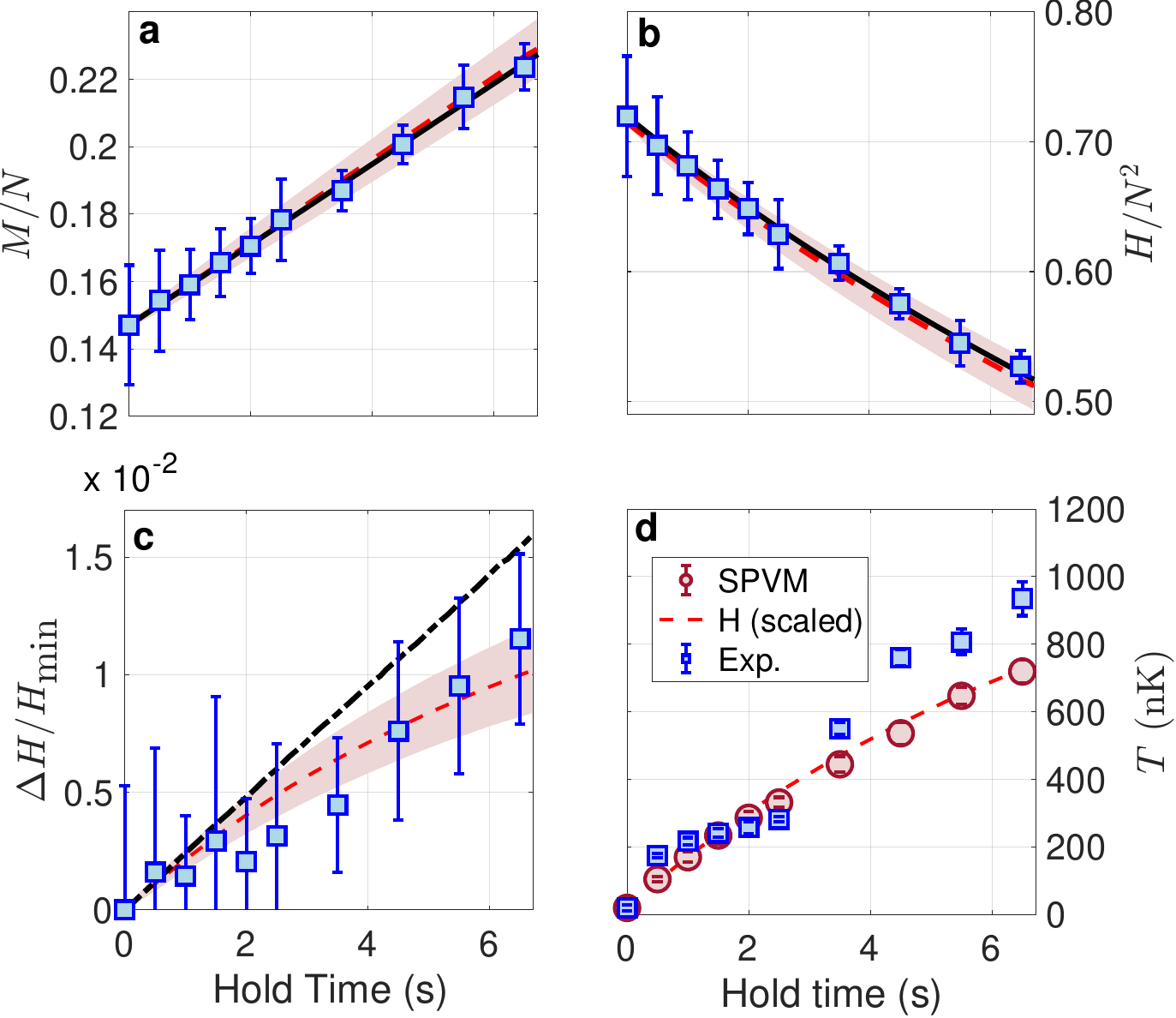}
    \caption{ \textbf{a} Angular momentum, and \textbf{b} energy, \textbf{c} percentage excess energy $\Delta H/H_\textrm{min}$, and \textbf{d} vortex temperature vs.~hold time. Experimental data points are shown with blue squares, while the solid black lines shown in \textbf{a} and \textbf{b} are fits to the scaling theory. Dashed red lines are simulations of Eq.~(\ref{eqn:StochasticPV}), for $\alpha = 3.25\times10^{-3}$, $\eta = 1.44\times10^{-3}$. Shaded areas and error bars show the standard deviation of the PV and experimental results respectively. A pure Weiner noise process corresponding to $\eta = 1.44\times10^{-3}$ is shown with a dash-dotted line in \textbf{c}. The dashed line in \textbf{d} is a rescaling of the SPVM excess energy to $\Delta H/k_BC$.}
    \label{fig:energy}
\end{figure}

Since the expansion is dominated by friction, this provides a means to determine $\alpha$;  we fit the scaling theory   to $H(t)$ and $M(t)$, with $\Omega(0) =  2\pi\times1.95$~Hz, shown in Fig.~\ref{fig:energy}\textbf{a},\textbf{b}. The two independent fits result in $\alpha = \{3.3(1), 3.2(1)\}\times 10^{-3}$ for $M(t)$ and $H(t)$ respectively.  To determine the heating rate characterised by $\eta$, we assess the energy in terms of the rescaled the coordinates $\zeta_j(t)$, which defines the excess energy, $\Delta H = H-H_{\rm min}$, of the OCP above the Wigner crystal energy $H_{\rm min}$. In these rescaled coordinates we observe growth in the energy $\Delta H_\textrm{exp}(t)$ by $\sim 1\%$, shown in Fig.~\ref{fig:energy}\textbf{c}. 

Starting from the initial experimental vortex positions, we compare the experimental results with $10^4$ numerical SPVM realizations of the dynamics from Eq.~\ref{eqn:StochasticPV},  with $\alpha = 3.25\times10^{-3}$. We find that $\eta = 1.44\times10^{-3}\, \xi^2/t_{\xi}$ minimizes the error between $\Delta H_\textrm{exp}(t)$ and $\Delta H_\textrm{SPVM}(t)$, where $\xi\sim 0.5~\mu$m is the healing length and $t_{\xi}\sim 0.5$~ms is the healing time. The cooling influence of the friction can be seen by comparing with a pure Weiner noise process, $\textrm{d}z_j = \sqrt{2 \eta}\, \mathrm{d} W_j$, which is shown with the dash-dotted line in Fig.~\ref{fig:energy}\textbf{c} and is seen to lead to approximately linear growth in $\Delta H$. Further support for the equilibrium melting scenario is provided by investigating the relation between $\Delta H$ and the vortex temperature $T$, which for OCP equilibria obeys $\Delta H = k_B C \Delta T$ for constant $C$~\cite{mazars2015melting}. Fig.~\ref{fig:energy}\textbf{d} shows excellent agreement between $T$ as determined from Monte Carlo calculations, and $\Delta H/k_B C$  as determined by dynamical simulations with $C \sim 5.9\times10^{-4}$.



The model Eq.~(\ref{eqn:StochasticPV}) was introduced phenomenologically in Ref.~\cite{reeves2022turbulent}, and a microscopically justified derivation from the stochastic projected Gross-Pitaevskii equation, a finite temperature reservoir theory for the weakly interacting Bose gas, was subsequently presented in Ref.~\cite{mehdi2023mutual}. The value predicted by this theory is $\eta = \alpha k_B T/(h n_0) \sim 1\times 10^{-5}\,\xi^2/t_{\xi}$, where $n_0$ is the 2D atom number density. This is $\sim 100\times$ smaller than the value for $\eta$ observed in our experiment. Since the condensate is in a hard-walled disc trap, we investigated whether vibrations of the walls can induce broad-band phonon excitations~\cite{navon2016emergence}. Using a quadrant photodetector, we find that the walls of the trap potential exhibit vibrations of $\sim \pm 0.5~\mu$m amplitude on the order of $10-140$~Hz. Gross-Pitaevskii (GPE) simulations suggest that such vibrations cause significant heating, and these vibrations are likely the cause of the increased value of $\eta$ relative to the prediction of Ref.~\cite{mehdi2023mutual}. This effect can be considered a feature of the experiment, allowing the observation of the melting transition.

\textit{Droplet Edge Characteristics ---} Finally, we turn to assessing recent theoretical predictions concerning the edge of the vortex matter droplet.  The low temperature states of the vortex matter form a discretized analog of a Rankine vortex, which in a classical fluid would have continuous and uniform vorticity $2\Omega = \Gamma N /2 \pi R_0$ for $r < R_0$ (and 0 otherwise). However, as shown by Bogatskiy and Wiegmann~\cite{bogatskiy2019edge},  anomalous stresses arising from the discreteness of the quantum vortex matter produce an irreducible topological fingerprint on the vortex distribution.  Within the liquid phase, the edge of the vortex matter is characterized by an excess of vortex density, the overshoot, with decaying oscillations into the bulk [see Fig.~\ref{fig:schematic}\textbf{b}]. 
The properties of the overshoot have garnered recent interest due to their connections to edge states in the FQHE~\cite{bogatskiy2019edge,cardoso2020boundary}. The overshoot is predicted to persist regardless of the number of particles, with the edge exhibiting a nonvanishing dipole moment $\bar d = 1/8\pi$~\cite{bogatskiy2019edge}. The vortex droplet is also squeezed by the anomolous stresses, yielding the radius $R = R_0 -h$, with $h$ dependent on the inter-vortex distance, $h = \ell /\sqrt{8 \pi}$.~\cite{bogatskiy2019edge}. 

\begin{figure}[t!]
    \centering
    \includegraphics[width = \columnwidth]{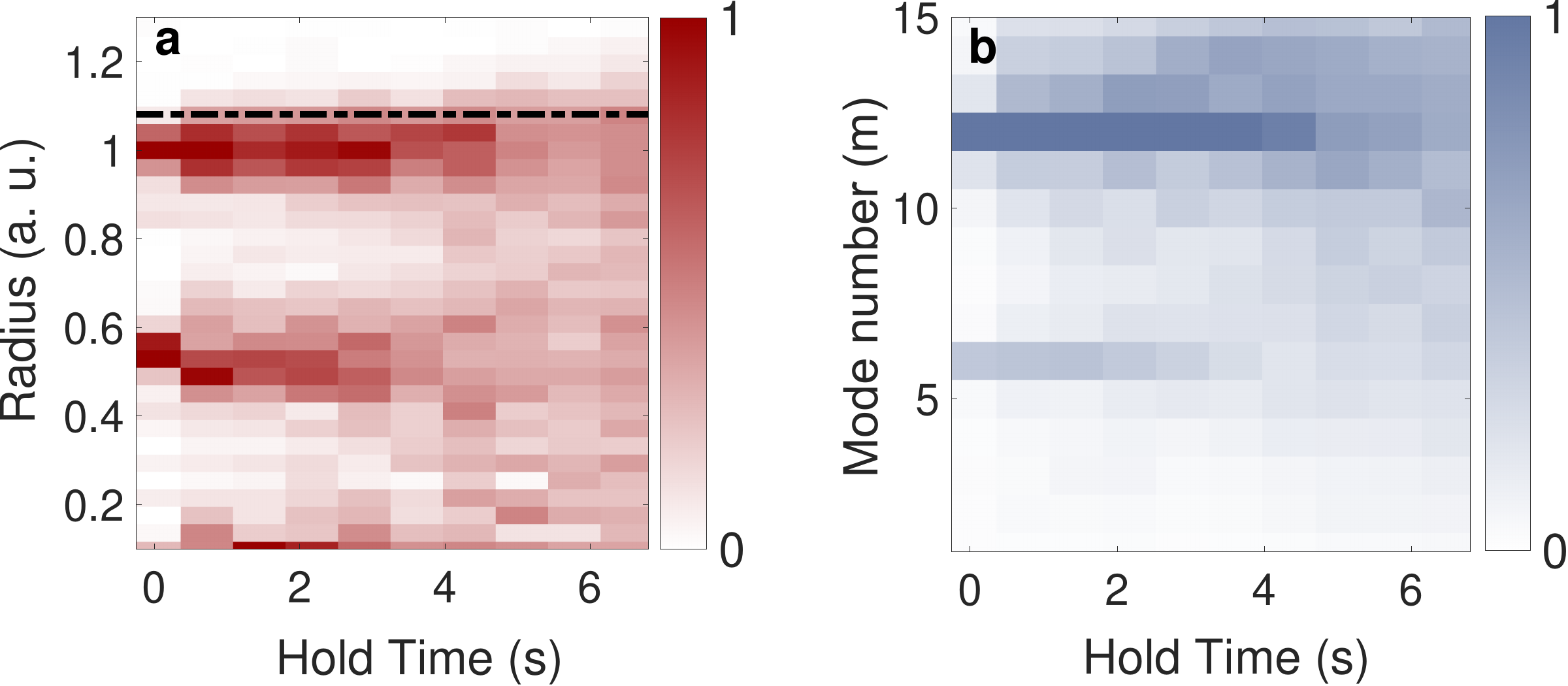}
    \caption{\textbf{a} Experimental vortex radius histograms in the rescaled coordinates. Both the edge density overshoot, indicated by the excess density at $r'=1$ throughout the hold times, and the loss of the internal lattice structure are clearly visible. The dashed line indicates the Rankine vortex radius $(1+\overline{d})$ -- the peak density remains below this line for the entire evolution, demonstrating the spatial squeezing of the cluster. \textbf{b} Normalized angular structure factor $s_m(t)$ is plotted as a function of hold time for the experimental data revealing evidence of crystallization at the cluster edge (see text).}
    \label{fig:combined_histograms}
\end{figure}

Signatures of the persistent edge overshoot can clearly be seen in the histogram data presented in Fig.~\ref{fig:lattice_histograms} at $t \geq 3.5$ s. 
The melting dynamics are further visualized in Fig.~\ref{fig:combined_histograms}, where radial histograms $\rho_i(r'(t))$ in the rescaled coordinates are combined for all ten hold times studied. The vertical axis is normalized so that $r'=1$ corresponds to $r = 23~\mu$m, the  position of the outermost vortices in the initial configuration. The edge overshoot is seen in the excess density at $r'=1$ persisting through longer hold times, while the internal crystal structure of the cluster disappears. The spatial compression of the cluster can also be analyzed as in Fig.~\ref{fig:combined_histograms}. The compression parameter is $\overline{d} = \ell/\sqrt{8 \pi}$, where $\ell = \rho^{-1/2}$, the inter-vortex spacing in the lattice. Setting $r'=1$ gives $\overline{d}\sim0.08$. The approximate Rankine vortex radius is thus indicated by the dashed line at $r' = 1.08$ in Fig.~\ref{fig:combined_histograms}. The peak of the experimental histograms remain within this limit throughout the temporal evolution.

Cardoso \emph{et al.}~\cite{cardoso2020boundary} have also recently argued that the edge overshoot manifests due to the crystal melting within the interior while the boundary remains crystalline.  To test this ``freezing at the edge" picture, we analysed our data for evidence of edge crystallization. We compute the normalized angular structure factor~\cite{hernandez2023universality}  $s_m(t) = (1/{N^2})\sum_{j=1}^{N}\sum_{k=1}^{N}e^{im[\theta_j(t)-\theta_k(t)]}$, where $\theta_j(t),\theta_k(t)$ are the angles of the $j$-th and $k$-th vortices relative to the cluster center. The structure function depends on the angle of the vortices only, and is strongly peaked when the vortices are regularly spaced at angles $\theta_m = 2\pi/m$. Thus, for the initial lattice configuration at time $t = 0$~s, the function is peaked for $m=6$ and $m=12$, corresponding to the inner and outer rings of vortices in the lattice. For increasing hold time the inner crystal structure is lost as the lattice melts, indicating that the angular positions of the innermost vortices are similarly randomized. However, the structure function remains strongly peaked around $m=12$ for much of the evolution, suggesting that crystalline structure on the boundary of the cluster is indeed maintained for longer~\cite{cardoso2020boundary}.

\textit{Discussion and outlook --- }
In summary we have experimentally realized a superfluid vortex simulator of the 2D OCP which demonstrates the equilibrium melting transition of the Wigner crystal. The simulator extends new controls for configurable initialization of OCP states and controlled heating via the superfluid phonon bath. The system is free of the defects present in type-II superconductors that affect vortex lattice melting dynamics~\cite{guillamon2009direct,roy2019melting,chandran2004domain}. In benchmarking our simulator we observe key features of the OCP and  melting transition, including excess density at the cluster edge, spatial compression of the cluster, and persistent crystallization on the cluster edge. Our experimental results show excellent agreement with a point-vortex model that accounts for the friction and noise, and the coefficients for these dissipative processes, retrieved via our scaling theory, are tightly constrained by the observed melting dynamics. Further study of the Wigner crystal melting would provide further insights into the dissipative terms contributing to vortex dynamics in quantum fluids, which is still an open question~\cite{sergeev2023mutual}. For example, there are generally two mutual friction coefficients, $\alpha,\alpha'$, as noted in superfluid Fermi gases and superfluid He-II. However, our measurements are consistent with $\alpha'\sim 0$ in the Bose gas. 

Our work further cements the utility of superfluids for exploring vortex matter. Additional scenarios, such as varying the initial vortex number and energy, are described in the Supplemental Materials~\cite{supp}. Beyond the equilibrium properties of the OCP, future experiments could test recent predictions of the anomalous hydrodynamics of vortex matter,  such as topological edge solitons~\cite{bogatskiy2019edge}, or other chiral surface modes~\cite{jeevanesan2022surface}. The experiment also establishes superfluid films as a platform for experimentally exploring melting transition dynamics, potentially addressing the existence of the intermediate hexatic phase~\cite{nosenko20092d,ryzhov2017berezinskii} in a background-defect-free system with logarithmic interactions, and whether they are driven by lattice-defect-induced quantum melting~\cite{nguyen2020fracton}. Such studies would benefit from a significantly increased vortex number, that may be realized in superfluid helium films~\cite{sachkou2019coherent}. The high level of control also suggests a programmatic approach to the systematic study of Tkachenko modes of the lattice~\cite{baym2003tkachenko,schweikhard2004rapidly}. In the present configuration the systematic heating may be problematic as one would like to maintain the system at fixed temperature in the liquid phase. This can be addressed by softening the edge of the disc trap with the projected DMD pattern and/or reducing the transduction of trap motion into the BEC as we have demonstrated in GPE simulations. 

We note recent related work on fluid states of vortices in rapidly rotating BECs~\cite{sharma2024thermal} that appeared during the preparation of this manuscript.

\textit{Acknowledgements --- }We thank Z. Mehdi, S. Szigetti, A. Bradley, and H. Rubinsztein-Dunlop for useful discussions. T.W.N. acknowledges the support of an Australian Research Council Future Fellowship FT190100306. M.T.R. acknowledges the support of an Australian Research Council Discovery Early Career Research Award DE220101548. Funding was also provided by the ARC Centre of Excellence for Engineered Quantum Systems (project number CE1101013). This research was also partially supported by the Australian Research Council Centre of Excellence in Future Low-Energy Electronics Technologies (project number CE170100039) and funded by the Australian Government. 

%

\section*{Methods}

\subsection{Experimental Methods}

A superfluid $^{87}$Rb Bose-Einstein condensate (BEC) is formed in a magnetically levitated optical trap produced by combining a focused Gaussian beam that provides vertical confinement crossed with a blue-detuned (repulsive) trapping potential derived from the direct projection of a digital micromirror device (DMD)~\cite{gauthier2016direct}. The resulting highly anisotropic BEC results in effective 2D vortex motion~\cite{gauthier2019giant,reeves2022turbulent,gauthier2019quantitative}. The DMD provides both the overall trapping potential and the dynamically movable pinning beams used to generate the vortices. \textit{In situ} images of the stirring sequence used to realize the cluster are shown in Fig.~\ref{fig:schematic}(b)~\footnote{See the Supplemental Materials~\cite{supp} for a video of the stirring sequence.}. The beams simultaneously spiral into the condensate by rotating at a constant rate of $1.2$~Hz and decreasing their radial position linearly in time, resulting in a peak speed of 250~$\mu$m/s, $\sim 0.2c$ at entry into the BEC, where $c$ the speed of sound. These create ``vortex pairs" on entry to the condensate in analogy with the ``chopsticks method"~\cite{samson2016deterministic}, one which becomes attached to the stirrer, and one which is virtual outside the condensate. The resulting positions of the vortices are determined by calculating the minimum energy configuration of Eq.~(\ref{eqn:PVHamiltonian}) for a $12~\mu$m lattice spacing~\footnote{corrections to $H$ due to the system boundary were found to be negligible for all the states considered}. On reaching the final positions, the angular rotation speed of the pinned vortices is accelerated to the solid-body rotation frequency $\Omega(0) =  2\pi\times1.95$~Hz over 200~ms.

After placing the vortices at their equilibrium positions in the rotating frame, they are released into the condensate by reducing the radius of the $5~\mu$m pinning beams to zero over 30~ms. We then track vortex positions over time by destructively detecting the vortices after a short $5~$ms time-of-flight (TOF), as shown in Fig.~\ref{fig:schematic}(b).  Vortex positions are displayed in the top row of Fig.~\ref{fig:lattice_histograms}, where each histogram contains $\sim 120$ individual realizations of the experiment, corresponding to  $\sim 2200$ vortices.

\subsection{Neural net vortex detection}
\begin{figure}
    \centering
    \includegraphics[width = \columnwidth]{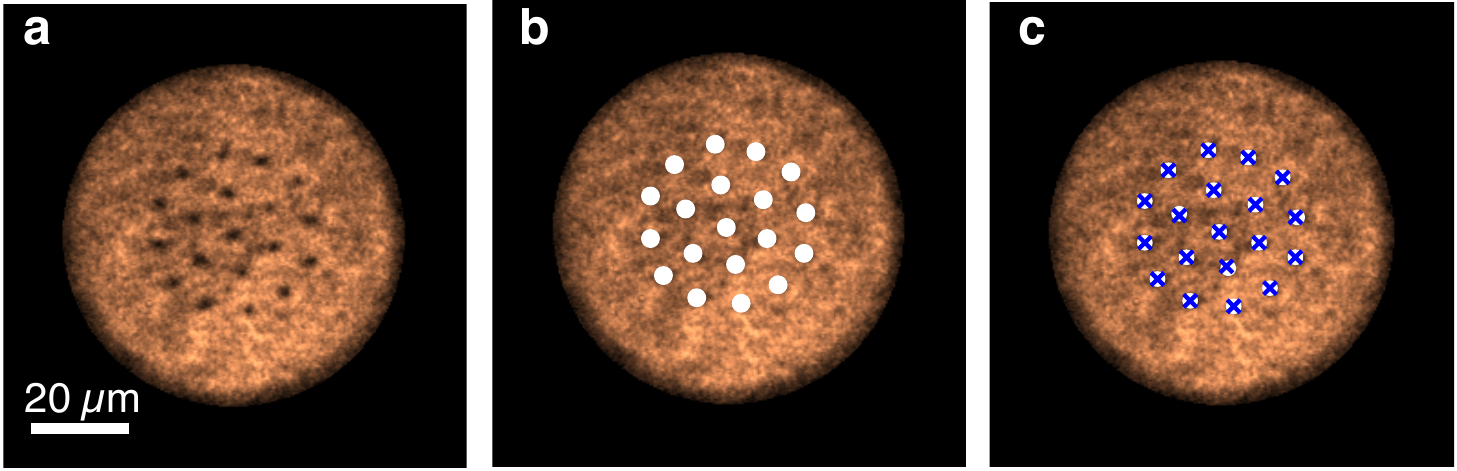}
    \caption{\textbf{a} Typical image of the BEC at $t_h=0$ after $5$~ms TOF showing vortices as dark density dips. \textbf{b} The resulting Gaussian blob based detection (white circles) used as the true position set. 
    \textbf{c} Detected vortex positions using the algorithm of Ref.~\cite{metz2021deep} (blue crosses) overlaid on the true positions (white circles).}
    \label{fig:detection}
\end{figure}

Vortices, seen as dark density dips in Fig.~\ref{fig:detection}\textbf{a}, are typically detected via an automated algorithm such as Gaussian blob detection~\cite{gauthier2019giant,rakonjac2016measuring}. Recently deep-learning image processing approaches were demonstrated for detecting vortices in numerically simulated BECs~\cite{metz2021deep}. We adapt the freely available code of Ref.~\cite{metz2021deep} to detect the vortices in the experimental data. The procedure is as follows. We first select a region of interest consisting of $256\times256$ pixels centered on the BEC. We then select a $128\times128$ pixel subgrid on which the vortices positions will be detected. The algorithm is trained on a subset of 50 images, where 40 images in the subset are used for training, and 10 of the images are used for testing. The true vortex positions for the 50 image set are determined via Gaussian blob detection, with the results manually checked for spurious detection/non-detection. We find that the deep-learning approach outperforms the Gaussian blob detection. For a data set of 122 images, corresponding to $t_h=0$, the results of the deep-learning detection are $\overline{N}_v = 18.7$ per image, with standard deviation $\sigma = 0.91$, and total detected vortices $N_v = 2284$. For the Gaussian blob algorithm, we find: $\overline{N}_v = 18.66$; $\sigma = 1.58$; $N_v = 2277$. Examining the results, the significantly larger standard deviation from the blob detection, while indicating greater shot-to-shot variation, also corresponds to increased numbers of spurious vortex detections away from the pinning beams. We therefore used the deep-learning method for the data presented in the main text.

The mean detected vortex number $\overline{N} =18.7$ can be compared to the programmed vortex number of $N=19$. Visual examination of the experimental runs reveals that additional/lost vortices can result from either shot-to-shot variation in stirring or issues in detection. For the 2D histogram data in the main text the full data sets are presented. However, when calculating the energy, angular momentum, and temperature we post-select the data to include only runs with $N=19$ detected vortices in the analysis as these quantities are strongly dependent on the vortex number. Approximately 60\% of the experimental runs contain 19 detected vortices, with most of the variation occurring due to detection variation rather than variation in the initial vortex creation process. The initial time step is particularly sensitive to the preparation and detection as $\Gamma$ changes rapidly for small variations in the initial vortex positions. For this time step we discard runs where the vortex positions vary significantly from the programmed positions which otherwise broaden the radial vortex histogram, as an estimate of the maximum value of $\Gamma$ typically achieved with the stirring process. The post-selection gives $87/122 \sim 71\%$ valid runs for the first time step, resulting in a $\Gamma \sim 230.5$ from the Monte Carlo fitting as presented in the main text. If we omit this post-selection, we find $\Gamma \sim 94$, within the liquid phase. The heating is sufficiently rapid that we find $\Gamma \sim 25.8$ at $t_h = 500$~ms and for $t >0$ the distributions are nearly unaffected by post selection of the data.

\subsection{Analysis of Friction and Noise}

We consider the case where the vortices are distributed with uniform density $\rho$. This cluster has a radius $R$ and rotates at angular frequency $\Omega$ given by
\begin{align}
R &= \sqrt{\frac{N}{\pi \rho}}, & \Omega = \frac{N \Gamma}{2 \pi R^2}.
\end{align}
The Hamiltonian is 
\begin{equation}
\mathcal{H} = H - \Omega M = - \sum_{i\neq j} \textrm{ln} {|z_i - z_j|}  + \Omega \sum_j
 |z_j|^2.
 \end{equation}
  The velocities in the rotating frame follow from  Hamilton's equations
 $\dot z_j = - i \partial \mathcal{H}/\partial z_j^*$, giving
 \begin{align}
\dot{z}_j& = u_j + v_j , 
\label{eqn:hamiltonDynamics}
\end{align}
where
\begin{align}
u_j & =- i \Omega z_j, &
 v_j &=   i \sum_{k\neq j} \frac{1}{(z_j - z_k)^*}
\end{align}
for which the ground state satisfies $u_j + v_j = 0$.

In an atomic superfluid the dynamics of quantized vortices have been shown to be well-described by a dissipative version of the Kirchoff equations including friction and noise~\cite{reeves2022turbulent,mehdi2023mutual}
\begin{equation}
    \mathrm{d}  z_j = (1 - i \alpha)v_j \,\mathrm{d} t + \sqrt{2 \eta}\, \mathrm{d} W_j.
    \label{eqn:dampedPVM}
\end{equation}
With $v_j$ defined as per Eq.~(\ref{eqn:hamiltonDynamics}), $\alpha$ being the mutual friction coefficient and $\eta$ the vortex diffusivity. Note that here there is no rotation term, as the equation describes the dynamics within the lab frame and in the experiment the thermal cloud is stationary. Under conservative evolution ($\alpha = \eta = 0$), the Wigner crystal evolves as 
\begin{equation}
z_j(t) = z_j(0) \textrm{exp}[i \Omega t].
\end{equation}
The friction is orthogonal to the conservative dynamics at all times and thus acts only radially on the Wigner crystal; under purely frictional motion the crystal therefore evolves according to 
\begin{equation}
z_j(t) = z_j(0) \textrm{exp}\left[\alpha   \int_0^t \Omega (\tau) d\tau \right].
\end{equation}
As the Kirchoff equations are invariant under the transformation $\{x,u,t\} \rightarrow \{\lambda x,\lambda^{-1}u,\lambda^2 t \}$, the above must satisfy a scaling equation of the form
\begin{align}
    \frac{z_j(t)}{z_j(0)}&=  \lambda(t), & \frac{\Omega(t)}{\Omega(0)} = \lambda(t)^{-2}. 
\end{align}
 Inserting this ansatz and solving for $\lambda$ yields the  scaling solution
 \begin{equation}
     \lambda(t) = (1 + 2\Omega(0) \alpha t )^{1/2},
 \end{equation}
where we have simplified using the property $\lambda(0) =1$ by definition. We note this result was previously postulated by appealing to a vortex fluid theory~\cite{stockdale2020universal} applicable for $N\gg 1$; here we see the result is exact for a Wigner crystal state of any $N$ (aside from the trivial case $N=1$). The dynamics of the Wigner crystal is thus stationary when viewed in terms of the scaled coordinates 
\begin{equation}
\zeta_j(t) =  \frac{z_j(t)}{\lambda(t)} \exp{\left[-i  \int_0^t \Omega(\tau) d\tau\right]} 
= \frac{z_j(t)}{\lambda(t)^{(1 +{i}/{2\alpha})}}.
\end{equation}

The relative coordinates $\zeta_j(t)$ determine the internal structure of the vortex matter and its statistical mechanical properties, whereas $\lambda(t)$ and $\Omega(t)$ encompass the global expansion and rotation in the dynamics. Under frictional dynamics the rescaled coordinates $\zeta_j$ tend towards the Wigner crystal equilibrium of the Hamiltonian~\cite{stockdale2020universal}. 

\begin{figure}[t]
    \centering
    \includegraphics[width = \columnwidth]{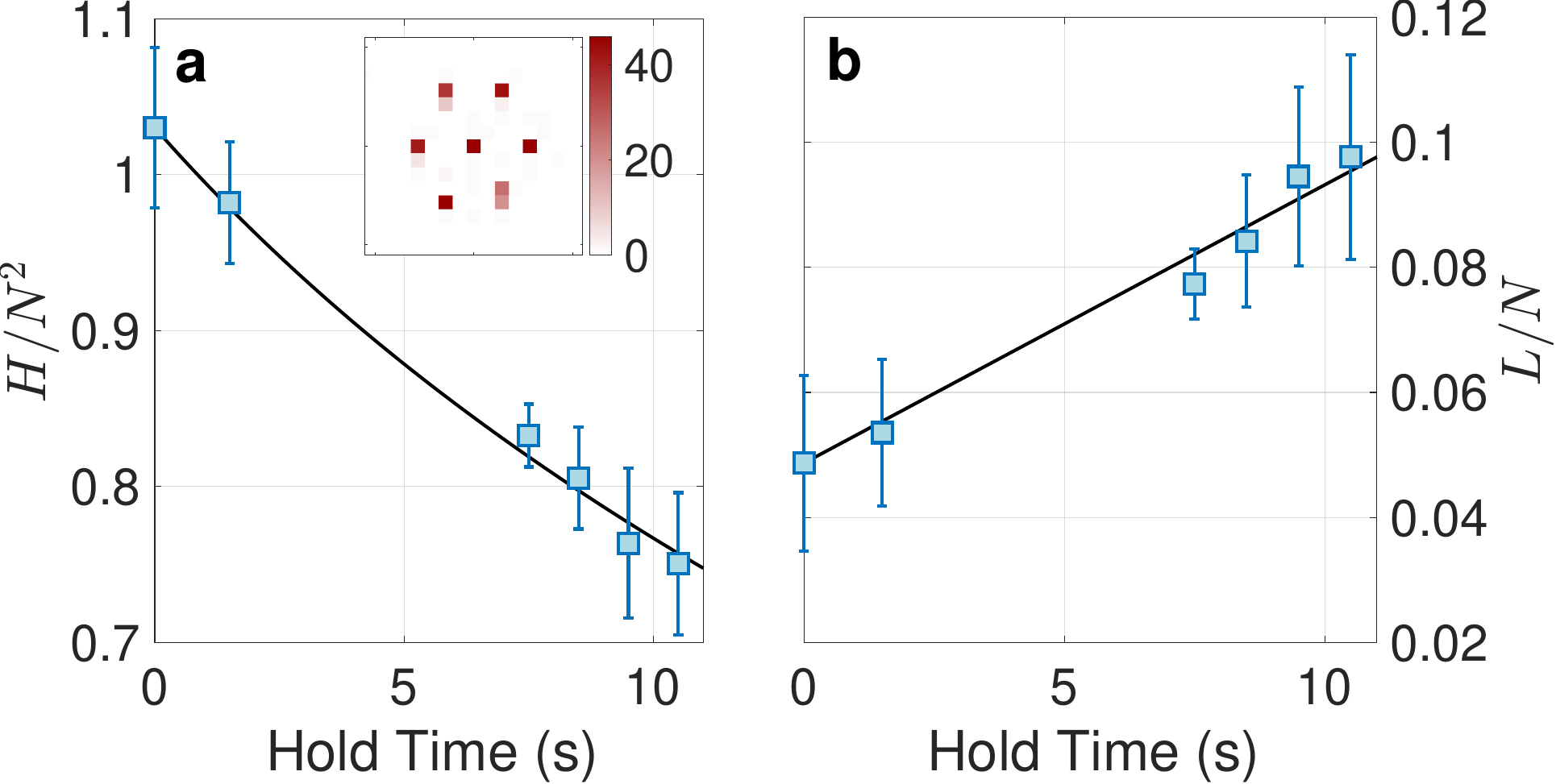}
    \caption{Scaling theory fits to a seven vortex initial ensemble (inset) with a $12~\mu$m lattice spacing. \textbf{a} Energy and \textbf{b} angular momentum fits giving $\alpha = \{3.4(3), 3.6(4)\}\times 10^{-3}$ respectively, consistent with the values for the 19-vortex lattice.}
    \label{fig:seven_vortex_dynamics}
\end{figure}

Returning now to the case when both friction and noise are present, we note that in the long time limit time limit the velocity  $v_j(t)\sim \lambda(t)^{-1}$ tends to zero due to frictional expansion, and the noise term $\mathrm{d}{W}$ thus dominates. The dynamics therefore tend towards a purely Brownian motion
\begin{equation}
  \lim_{t \rightarrow \infty}  \mathrm{d}z_j =\sqrt{2\eta}dW_j, 
\end{equation}
for which the distribution tends to a Gaussian, irrespective of the initial conditions. We note that under Hamiltonian dynamics the Gaussian distribution of vortices corresponds to one of infinite temperature in the mean-field approximation~\cite{reeves2022turbulent}. The vortex matter is thus always heated to an infinte temperature state in the long time limit, when evolving under the influence of a stationary thermal cloud.

It is worth remarking that the situation changes considerably if the friction associated with the thermal cloud damps the vortex matter to the rotating frame; in this case in one must make the replacement $v_j \rightarrow u_j+v_j$ in Eq.~(\ref{eqn:dampedPVM}), and the friction drives the vortex matter towards the ground state rotating at $\Omega$. In this situation detailed balance between dissipation and fluctuations can be achieved.

In order to further verify the scaling theory solution and its $N$-independence we experimentally investigated the damping of a seven-vortex lattice, see Fig.~\ref{fig:seven_vortex_dynamics}. The lattice spacing is similarly chosen to be $12~\mu$m identical to the lattice investigated in the main text. We again obtain an excellent fit with the scaling theory, $\alpha = \{3.4(3), 3.6(4)\}\times 10^{-3}$, consistent with the value of $\alpha$ obtained for the 19-vortex lattice within the uncertainty.

\clearpage

\section*{Supplemental Materials}
\setcounter{subsection}{0}
\subsection{Conservative dynamics testing of ergodicity} 

\begin{figure}[b]
    \centering
    \includegraphics[width = 0.9\columnwidth]{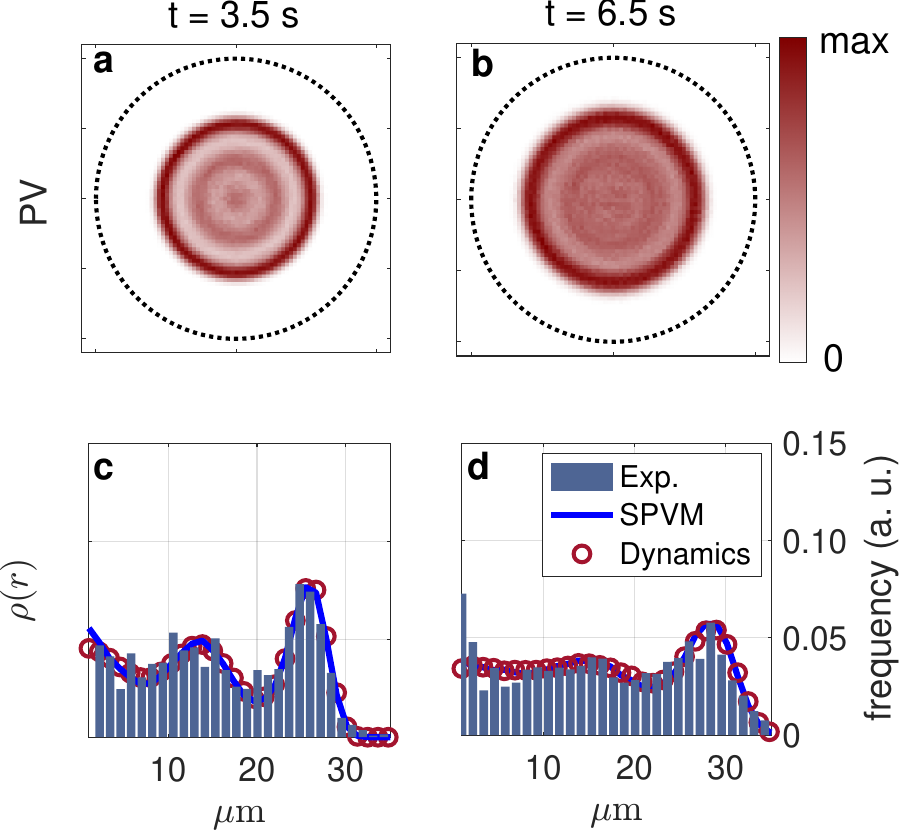}
    \caption{Dynamics test of ergodicity of the PV system. 2D histograms of the vortex trajectories are shown in the top row, while the radial experimental data (bars), SPVM modeling (solid blue line), and single-trajectory dynamics (open red circles) are shown in the bottom row.}
    \label{fig:dynamics_fig}
\end{figure} 

While the ensemble averages shown in Fig.~\ref{fig:lattice_histograms} and Fig.~\ref{fig:energy} of the main text exhibit variance in energy and angular momentum at later times we anticipate that the mean values are representative of the microcanonical ensemble for a given energy and angular momentum. To test this hypothesis we numerically simulate conservative dynamics, $\mathrm{d}  z_j = v_j \,\mathrm{d} t$, over $\sim 323$~s. We begin from a sample of the stochastic point vortex model Eq.~(\ref{eqn:StochasticPV}) data ensemble that closely matches the mean energy and angular momentum at a given hold time. Two-dimensional and radial histograms of these single trajectories are shown in Fig.~\ref{fig:dynamics_fig}. While we only show samples at $t_h = 3.5$~s and $t_h = 6.5$~s, we find that a close match between the dynamics and the ensemble average at all hold times. These results further support our conclusion that the mean values are representative of thermal quasi-equilibrium states of the vortex system.

\subsection{Vortex position shift induced melting}
\begin{figure}[b]
    \centering
    \includegraphics[width = \columnwidth]{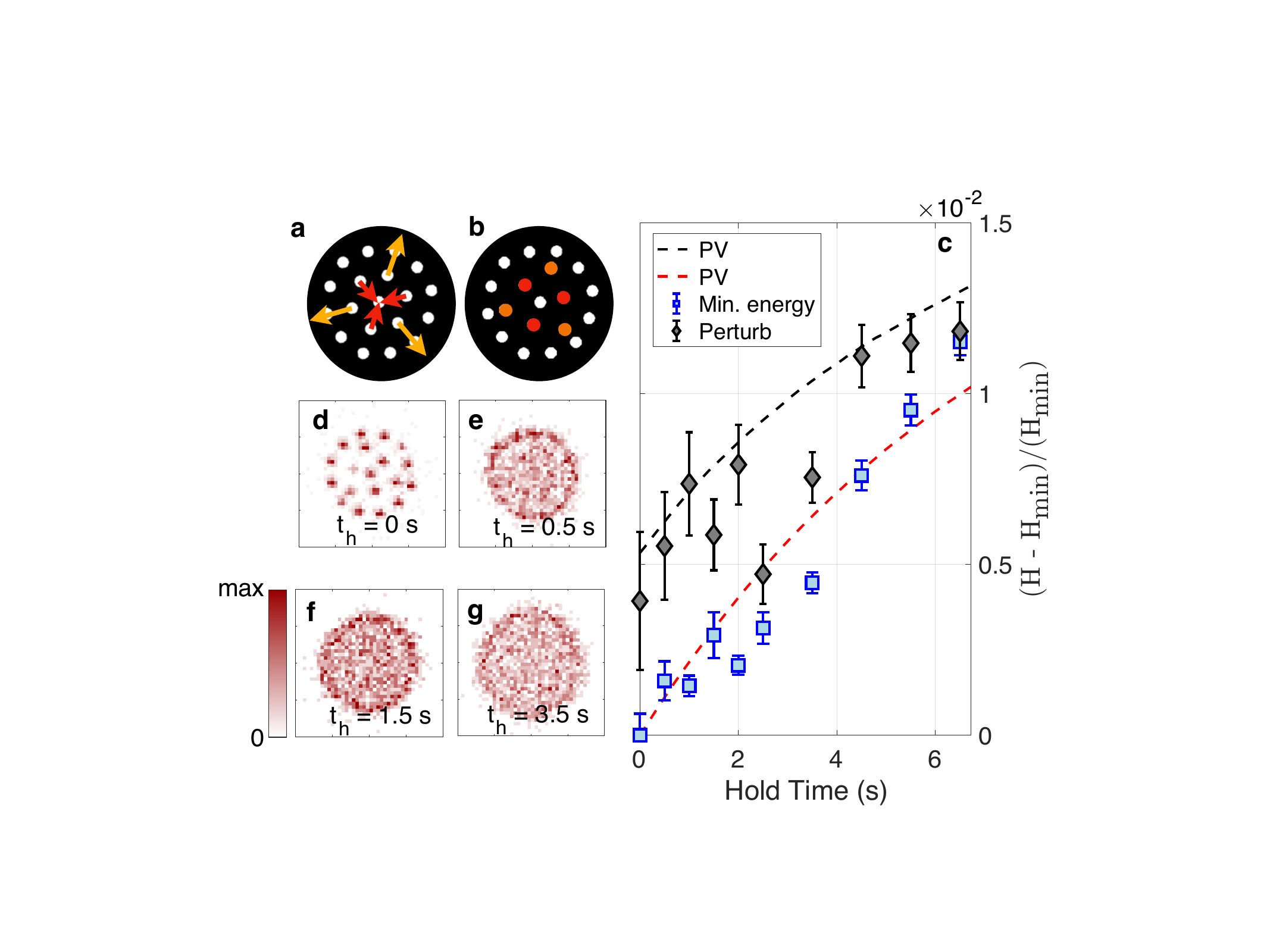}
    \caption{Melting dynamics starting from a perturbed initial condition. The inner ring of vortices are alternately shifted in and out as compared to the minimum energy state \textbf{a},\textbf{b}. ~\textbf{c} The result is an increase in the initial energy resulting in a rapid equilibration to larger early temperature. The experimental data are shown with blue squares and gray diamonds for the minimum energy (main text) and perturbed initial configurations respectively. Dashed lines are PV simulations using the values of $\eta$ and $\alpha$ from the main text. \textbf{d}-\textbf{g} Experimental histograms of the vortex positions demonstrating that the edge density overshoot is observed earlier in the evolution. This is consistent with the increased energy at early times.}
    \label{fig:perturb}
\end{figure}

While we have focused on initializing the system to its minimum energy state Wigner crystal, the flexibility of the vortex chopsticks method means that the vortex positions can be adjusted arbitrarily to modify the initial energy of the cluster. In further experiments we shifted the initial position of the inner ring of 6 vortices by alternately moving the vortices in and out by $\pm 30\%$ of their radial positions, see Fig.~\ref{fig:perturb}\textbf{a},\textbf{b}. This results in a $\sim0.5\%$ increase of the initial energy relative to the minimum energy in the rescaled coordinates $z'$. Correspondingly, in the subsequent dynamics we observe near instantaneous loss of the initial structure and emergence of the edge density overshoot within $\sim 1$~s of hold time consistent with the increased initial energy of the system as shown in Fig.~\ref{fig:perturb}\textbf{d}-\textbf{g}. Over 6.5~s of dynamics the energy grows by a similar fraction as for the minimum energy case. We find SPVM vortex modeling is consistent with the experimental data for the same $\alpha = 3.25\times10^{-3}$ and $\eta = 1.44\times10^{-3}\xi^2/t_{\xi}$ used in the main text.

\subsection{Vortex pair correlation function}
The pair correlation function  characterizes the state of matter of a system with solid, liquid and gas states exhibiting long-, medium- and short-range order, respectively. We calculate the vortex pair correlation function 
\begin{equation}
    \hat{g}\left(r,t\right) = \left\langle\sum_{i=1}^N \sum_{j \neq n} \frac{\delta\left(r_{ij}(t)-r\right)}{2\theta r} \right\rangle,
    \label{eq:gHat}
\end{equation}
from the experimental data at each hold time, where $N$ is the number of detected vortices in the system, $r_{ij} = \left|z_i - z_j\right|$, and $\theta$ is the angular extent of a circle located at $r_i = \left|z_i\right|$ with radius $r$ inside the circular boundary $R=100~\mu$m. The $\theta$ factor, 

\begin{equation}
    \theta = \left\{
    \begin{array}{lc}
        \pi, & \text{if } r \leq R-r_i, \\
        \arctan\left(\frac{r^2+r_i^2-R^2}{\sqrt{2R^2\left(r_i^2+r^2\right)-R^4-\left(r^2-r_i^2\right)^2}}\right),  & \text{otherwise,} 
    \end{array}
\right. 
\end{equation}

\noindent corrects for the boundary of the finite circular system. We define a normalized pair correlation function by taking \eqnreft{eq:gHat} and integrating over bins $\Delta r$ with center located at $r_n = (n+1/2)\Delta r \hspace{3pt}\forall \hspace{3pt} n \in \mathbb{N}_0 \leq R/\Delta r-1/2$:

\begin{equation}
    g(r_n, t) = \frac{1}{\Delta r \int_0^{2R} dr\hat{g}\left(r,t\right)} \int_{r_n-\Delta r/2}^{r_n+\Delta r/2} dr\hat{g}\left(r,t\right).
\end{equation}

The initial values of $g(r, t)$ at $t=0$~s demonstrate long-range order, seen as sharp well-separated peaks as expected for a lattice in the solid phase for $t_h=0$~s and $t_h=0.5$~s. As the system evolves we observe melting via the correlation function -- the peaks and troughs attenuate but still remain visible at $t_h=2.5$~s. At the end of the evolution, $t_h=6.5$~s, the peak structure has fully attenuated (except for the initial overshoot) indicating intermediate range order. The transition from long-range to intermediate-range order provides further evidence of a solid-to-liquid transition in the vortex matter.

\begin{figure}[b]
    \centering
    \includegraphics[width = \columnwidth]{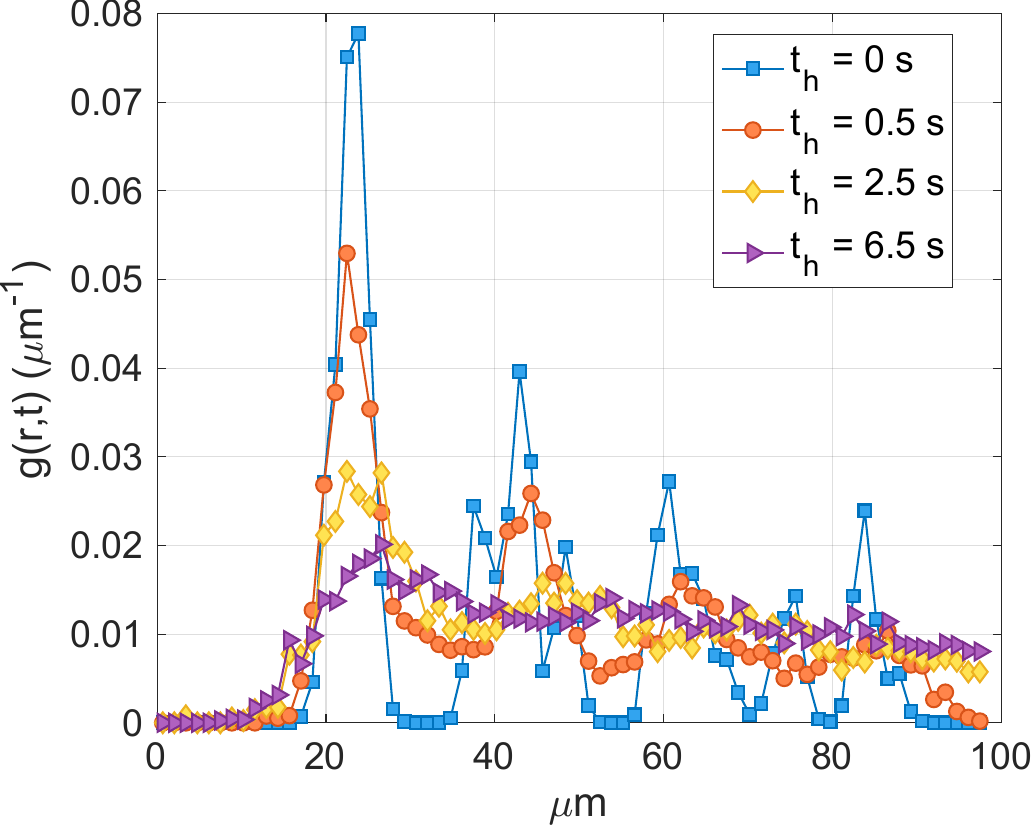}
    \caption{Pair correlation function of the vortex positions for hold times 0~s (blue squares), 0.5~s (orange circles), 2.5~s (yellow diamonds), and 6.5~s (purple triangles).}
    \label{fig:piarcorr}
\end{figure}

\end{document}